\documentclass[a4paper]{article}
\usepackage{RR}
\RRNo{6254}
\usepackage{hyperref}

\usepackage{amssymb}
\usepackage{amsmath}
\usepackage{graphicx}
\usepackage{epsfig}

\newtheorem{Theorem}{Theorem}
\newtheorem{Lemma}{Lemma}

\newtheorem{Proposition}{Proposition}

\RRdate{July 2007} 
\RRauthor{ Eitan Altman\thanks{INRIA Sophia Antipolis,
Altman@sophia.inria.fr} \and Konstantin Avrachenkov\thanks{INRIA
Sophia Antipolis, K.Avrachenkov@sophia.inria.fr} \and Andrey
Garnaev\thanks{St.Petersburg State University, agarnaev@rambler.ru}
}

\authorhead{Altman, Avrachenkov \& Garnaev}

\RRtitle{Solution analytique des jeux de water-filling
sym\'etriques}

\RRetitle{Closed form solutions for symmetric\\ water filling games}

\titlehead{Closed form solutions for symmetric water filling games}
\RRnote{This work was supported by BioNets European project and by
joint RFBR and NNSF Grant no.06-01-39005.}

\RRresume{Nous \'etudions le contr\^ole de puissance dans le cadre
de l'optimisation et dans celui de la th\'eorie des jeux. Dans le premier,
il y a un seul agent qui assigne les ressources du
r\'eseau tandis que dans le deuxi\`eme, les utilisateurs
se partagent les ressources du r\'eseau selon
l'\'equilibre de Nash. La
solution de ces probl\`emes est bas\'ee sur la m\'ethode du
water-filling. 
On calcule des multiplicateurs de Lagrange en utilisant une m\'ethode
de bisection pour resoudre des \'equations non lin\'eaires.
Nous fournissons ici une solution analytique
au probl\`eme du water-filling, qui nous permet de le r\'esoudre
en un nombre fini d'op\'erations. En outre, nous produisons une
solution analytique de l'\'equilibre de Nash dans le cadre de la th\'eorie 
des jeux.
Nous \'etudions un jeu sym\'etrique en terme d'interf\'erence avec un
nombre arbitraire d'utilisateurs. Quoique le jeu soit sym\'etrique,
il y a une structure hi\'erarchique induite par la quantit\'e des
ressources disponibles pour les utilisateurs. Nous utilisons cette
structure pour effectuer une r\'eduction successive
du jeu. En plus de son \'el\'eguence math\'ematique, la solution
analytique permet d'\'etudier des cas limites quand le coefficient
d'interf\'erence est petit ou grand. Nous fournissons une preuve
simple de la convergence de l'algorithme it\'eratif de
water-filling (l'algorithme de meilleur r\'eponse). Il s'av\`ere que la
convergence de l'agorithme est ralentie
quand le coefficient d'interf\'erence est proche de l'unit\'e. En
utilisant la solution analytique, nous pouvons \'eviter ce
probl\`eme. Aussi, nous comparons l'approche non coop\'erative \`a
l'approche coop\'erative et montrons que l'approche non
coop\'erative fournit une distribution des ressources plus \'equitable.}

\RRabstract{We study power control in optimization and game
frameworks. In the optimization framework there is a single decision
maker who assigns network resources and in the game framework users
share the network resources according to Nash equilibrium. The
solution of these problems is based on so-called water-filling
technique, which in turn uses bisection method for solution of
non-linear equations for Lagrange multiplies. Here we provide a
closed form solution to the water-filling problem, which allows us
to solve it in a finite number of operations. Also, we produce a
closed form solution for the Nash equilibrium in symmetric Gaussian
interference game with an arbitrary number of users. Even though the
game is symmetric, there is an intrinsic hierarchical structure
induced by the quantity of the resources available to the users. We
use this hierarchical structure to perform a successive reduction of
the game. In addition, to its mathematical beauty, the explicit
solution allows one to study limiting cases when the crosstalk
coefficient is either small or large. We provide an alternative
simple proof of the convergence of the Iterative Water Filling
Algorithm. Furthermore, it turns out that the convergence of
Iterative Water Filling Algorithm slows down when the crosstalk
coefficient is large. Using the closed form solution, we can avoid
this problem. Finally, we compare the non-cooperative approach with
the cooperative approach and show that the non-cooperative approach
results in a more fair resource distribution.}

\RRmotcle{r\'eseaux sans fils, contr\^ole de puissance, jeu
water-filling sym\'etrique, \'equilibre de Nash, co\^ut de l'anarchie}

\RRkeyword{wireless networks, power control, symmetric water-filling
game, Nash equilibrium, price of anarchy}

\RRprojets{MAESTRO} 
\RRtheme{\THCom} 
\URSophia 
\begin{document}
\makeRR

\section{Introduction}

In wireless networks and DSL access networks the total available
power for signal transmission has to be distributed among several
resources. In the context of wireless networks, the resources may
correspond to frequency bands (e.g. as in OFDM), or they may
correspond to capacity available at different time slots. In the
context of DSL access networks, the resources correspond to
available frequency tones. This spectrum of problems can be
considered in either optimization scenario or game scenario. The
optimization scenario leads to ``Water Filling Optimization
Problem'' \cite{CT1991,GV1997,TV2005} and the game scenario leads to
``Water Filling Game'' or ``Gaussian Interference Game''
\cite{ElG05,PR03,PPR04,Yu2002}. In the optimization scenario, one
needs to maximize a concave function (Shannon capacity) subject to
power constraints. The Lagrange multiplier corresponding to the
power constraint is determined by a non-linear equation. In the
previous works \cite{CT1991,GV1997,TV2005}, it was suggested to find
the Lagrange multiplier by means of a bisection algorithm, where
comes the name ``Water Filling Problem''. Here we show that the
Lagrange multiplier and hence the optimal solution of the water
filling problem can be found in explicit form with a finite number
of operations. In the multiuser context, one can view the problem in
either cooperative or non-cooperative setting. If a centralized
controller wants to maximize the sum of all users' rates, the
controller will face a non-convex optimization problem with many
local maxima \cite{SCGC02}. On the other hand, in the
non-cooperative setting, the power allocation problem becomes a game
problem where each user perceives the signals of the other users as
interference and maximizes a concave function of the noise to
interference ratio. A natural approach in the non-cooperative
setting is the application of the Iterative Water Filling Algorithm
(IWFA) \cite{YGC02}. Recently, the authors of \cite{LP06} proved the
convergence of IWFA under fairly general conditions. In the present
work we study the case of symmetric water filling game. There is an
intrinsic hierarchical structure induced by the quantity of the
resources available to the users. We use this hierarchical structure
to perform a successive reduction of the game, which allows us to
find Nash equilibrium in explicit form. In addition, to its
mathematical beauty, the explicit solution allows one to find the
Nash equilibrium in water filling game in a finite number of
operations and to study limiting cases when the crosstalk
coefficient is either small or large. As a by-product, we obtain an
alternative simple proof of the convergence of the Iterative Water
Filling Algorithm. Furthermore, it turns out that the convergence of
IWFA slows down when the crosstalk coefficient is large. Using the
closed form solution, we can avoid this problem. Finally, we compare
the non-cooperative approach with the cooperative approach and
conclude that the cost of anarchy is small in the case of small
crosstalk coefficients and that the the decentralized solution is
better than the centralized one with respect to fairness.
Applications that can mostly benefit from decentralized
non-cooperative power control are ad-hoc and sensor networks with no
predefined base stations \cite{HCB00, LG97, KG99}. An interested
reader can find more references on non-cooperative power control in
\cite{AAMP07,ElG05}. We would like to mention that the water filling
problem and jamming games with transmission costs have been analyzed
in \cite{AAG07}.

The paper is organized as follows: In Section~\ref{sec:SUsetup} we
recall the single decision maker setup of the water filling
optimization problem and provide its explicit solution. Then in
Sections~\ref{sec:MUsetup}-\ref{sec:GameClosedForm} we formulate
multiuser symmetric water filling game and characterize its Nash
equilibrium, also we give an alternative simple proof of the
convergence of the iterative water filling algorithm and suggest the
explicit form of the users' strategy in the Nash equilibrium. In
Section~\ref{sec:numeric} we confirm our finding with the help of
numerical examples and compare the decentralized approach with the
centralized one.

\section{Single decision maker}
\label{sec:SUsetup}

First let us consider the power allocation problem in the case of a
single decision maker. The single decision maker (also called
``user'' or ``transmitter'') wants to send information using $n$
independent resources so that to maximize the Shannon capacity. We
further assume that resource $i$ has a ``weight'' of $\pi_i$.

\noindent Possible interpretations:
\begin{itemize}
\item[(i)] The resources may correspond to capacity available at
different time slots; we assume that there is a varying environment
whose state changes among a finite set of states $i \in [1,n]$,
according to some ergodic stochastic process with stationary
distribution $\{\pi_i\}_{i=1}^n$. We assume that the user has
perfect knowledge of the environment state at the beginning of each
time slot.
\item[(ii)] The resources may correspond to frequency bands (e.g. as in
OFDM) where one should assign different power levels for different
sub-carriers \cite{TV2005}. In that case we may take $ \pi_i = 1/n$
for all $i$.
\end{itemize}

The strategy of user is $T=(T_1,\ldots,T_n)$ with $\sum_{i=1}^n
\pi_i T_i = \bar{T}$, $T_i\geq 0, \pi_i>0$ for $i\in [1,n]$ and
$\bar{T}>0$. As the payoff to user we take the Shannon capacity
$$
v(T) = \sum_{i=1}^n \pi_i\ln\left(1+T_i/N^{0}_{i}\right),
$$
where  $N^{0}_{i}>0$ is the noise level in the sub-carrier $i$.

We would like to emphasize that this generalized description of the
water-filling problem can be used for power allocation in time as
well as power allocation in space-frequency. Following the standard
water-filling approach \cite{CT1991,GV1997,TV2005} we have the
following result.

\begin{Theorem}
\label{thm001} Let $T_i(\omega)=\left[1/\omega-N^0_i\right]_{+}$ for
$i\in [1,n]$ and $H_T(\omega)=\sum_{i=1}^n \pi_i T_i(\omega)$. Then
$T(\omega^*)=(T_1(\omega^*),\ldots,T_n(\omega^*))$ is the unique
optimal strategy and its payoff is $v(T(\omega^*))$ where $\omega^*$
is the unique root of the equation
\begin{equation}
\label{eqnWF} H(\omega) = \bar{T}.
\end{equation}
\end{Theorem}

In the previous studies of the water-filling problems it was
suggested to use numerical (e.g., bisection) method to solve the
equation (\ref{eqnWF}). Here we propose an explicit form approach
for its solution.

\noindent Without loss of generality we can assume that
\begin{equation}
\label{eqnNOrder} 1/N^0_1\geq 1/N^0_2 \geq \ldots\geq 1/N^0_{n}.
\end{equation}
Then, since $H(\cdot)$ is decreasing, we have the following result:
\begin{Theorem}
\label{thm1player} The solution of the water-filling optimization
problem is given by
\begin{equation*}
\label{lblWF} T_i^* =
  \begin{cases}
    \Bigl(\displaystyle \bar T+\sum_{t=1}^k\pi_t(N^0_t-N^0_i)\Bigr)\Bigl/\Bigl(\displaystyle \sum_{t=1}^k \pi_t \Bigr), &  i \leq k, \\
    0, &  i>k,
  \end{cases}
\end{equation*}
where $k$ can be found from the following condition:
\begin{equation*}
\varphi_k<\bar T \leq \varphi_{k+1},
\end{equation*}
where
$$
\varphi_t=\sum_{i=1}^t\pi_i(N^0_t-N^0_i)\mbox{ for } t\in [1,n].
$$
\end{Theorem}

Thus, contrary to the numerical (bisection) approach, in order to
find an optimal resource allocation we need to execute only a {\it
finite} number of operations.

\section{Symmetric water filling game}
\label{sec:MUsetup}

Let us now consider a multi-user scenario. Specifically, we consider
$L$ users who try to send information through $n$ resources so that
to maximize their transmission rates. The strategy of user $j$ is
$T^j=(T^j_1,\ldots,T^j_n)$ subject to
\begin{equation}
\label{eqn00} \sum_{i=1}^n \pi_i T^j_i = \bar{T^j},
\end{equation}
\noindent where $\bar{T^j}>0$ for $j\in[1,L]$. The element $T^j_i$
is the power level used by transmitter $j$ when the environment is
in state $i$. The payoff to user $j$ is given as follows:
\begin{equation*}
v^j(T^1,\ldots,T^L) = \sum_{i=1}^n \pi_i
\ln\left(1+\frac{\alpha^j_{i} T^j_i}{N^{0}_i+g_i
\sum_{k\not=j}\alpha^k_{i} T^k_i}\right),
\end{equation*}
where $N^{0}_i$ is the noise level and  $g_i\in (0,1)$ and
$\alpha^j_{i}$ are fading channel gains of user $j$  when the
environment is in state $i$.  These payoffs correspond to Shannon
capacities. The constraint (\ref{eqn00}) corresponds to the average
power consumption constraint. This is an instance of the Water
Filling or Gaussian Interference Game
\cite{ElG05,PR03,PPR04,Yu2002,YGC02}. In the important particular
cases of OFDM wireless network and DSL access network, $\pi_i=1/n,
i=1,...,n$.

We will look for a Nash Equilibrium (NE) of this problem. The
strategies $T^{1*}$,\ldots,$T^{L*}$ constitute a NE, if for any
strategies $T^{1}$,\ldots,$T^{L}$ the following inequalities hold:
\begin{equation*}
\begin{split}
v^1(T^1, T^{2*},\ldots,T^{L*}) &\leq  v^1(T^{1*},T^{2*}\ldots,T^{L*}),\\
&\cdots\\
v^L(T^{1*},\ldots,T^{(L-1)*},T^{L}) &\leq
v^L(T^{1*},\ldots,T^{(L-1)*},T^{L*}).
\end{split}
\end{equation*}
For finding NE of such game usually the following numerical
algorithm is applied. First, a strategy of $L-1$ users (say, user
2,\ldots, $L$) are fixed. Then, the best reply of user 1 is found
solving the Water Filling optimization problem.  Then, the best
reply of user 2 on these strategies of the users is found solving
the optimization problem and so on. It is possible to prove that
under some assumption on fading channel gains this sequence of the
strategies converge to a NE \cite{LP06}.

In this work we restrict ourselves to the case of symmetric game
with equal crosstalk coefficients. This situation can for example
correspond to the scenario when the users are situated at about the
same distance from the base station. Namely, we assume that
$\alpha^1_{i}=\ldots=\alpha^L_{i}$ and $g_i=g$ for $i\in (0,1)$. So,
in our case the payoffs to users are given as follows
\begin{equation*}
v^j(T^1,\ldots,T^L) = \sum_{i=1}^n \pi_i \ln\left(1+\frac{
T^j_i}{N^{0}_{i}+g \sum_{k\not=j} T^k_i}\right),
\end{equation*}
where $N^{0}_{i}=N^{0}/\alpha_{i}$, $i\in [1,n]$ and without loss of
generality we can assume that the channels are arranged in such a
way that the inequalities (\ref{eqnNOrder}) hold. We would like to
emphasize that the dependance of $N^{0}_{i}$ on $i$ allows us to
model an environment with varying transmission conditions.

For this problem we propose a new algorithm of finding the NE. The
algorithm is based on closed form expressions and hence it requires
only a finite number of operations. Also, explaining this algorithm
we will prove that the game has the unique NE under assumption that
$g\in (0,1)$.

Since $v_j$ is concave on $T^j$, the Kuhn-Tucker Theorem  implies
the following theorem.
\begin{Theorem}
\label{thm02} $(T^{1*},\ldots, T^{L*})$ is a Nash equilibrium if and
only if there are non-negative $\omega^j$, $j\in [1,L]$ (Lagrange
multipliers) such that
\begin{equation}
\begin{split}
\label{eqn1} \frac{\partial}{\partial T^j_i}v^j(T^{1*},\ldots,
T^{L*})&=\frac{\displaystyle 1}{\displaystyle T^{j*}_i+
N^{0}_{i}+g\sum_{k\not=j}T^{k*}_i }\\
  &\begin{cases}
    =\omega^j& \text{for }  T^{j*}_i>0,\\
    \le \omega^j& \text{for } T^{j*}_i=0.
  \end{cases}
\end{split}
\end{equation}
\end{Theorem}
It is clear that all $\omega^j$  are positive.

The assumption that $g<1$  is crucial for uniqueness of equilibrium
as it is shown in the following proposition.
\begin{Proposition} \label{rmk1}
For $g=1$  the symmetric water filling game has infinite number
(continuum) of Nash equilibria.
\end{Proposition}

{\it Proof}. Suppose that $(T^{1*},\ldots, T^{L*})$ is a Nash
equilibrium. Then, by Theorem~\ref{thm02}, there are non-negative
$\omega^j$, $j\in [1,L]$  such that
\begin{equation*}
1\bigl/\bigl( N^{0}_{i}+\sum_{k=1}^LT^{k*}_i\bigr)
  \begin{cases}
    =\omega^j& \text{for }  T^{j*}_i>0,\\
    \le \omega^j& \text{for } T^{j*}_i=0.
  \end{cases}
\end{equation*}
Thus, $\omega^1=\ldots=\omega^L=\omega$. So, $T^{1*}_i$, \ldots,
$T^{L*}_i$, $i\in [1,n]$ have to be any non-negative such that
$$
\sum_{k=1}^L T^{k*}_i=\pi_i[1/\omega-N^0_i]_+,
$$
and
$$
\sum_{i=1}^n\pi_iT^{k*}_i=\bar T^k \mbox{ for } k\in [1,L],
$$
where $\omega$ is the unique positive root of the equation
$$
\sum_{i=1}^n [1/\omega-N^0_i]_+ = \sum_{k=1}^L \bar T^k.
$$
It is clear that there are infinite number of such strategies. For
example, if $T^{a*}_i$ and $T^{b*}_i$, $i\in [1,n]$ ($a\not=b$) is
the one of them and $T^{a*}_k, T^{b*}_k>0$ and $T^{a*}_k, T^{b*}_m>$
for some $k$ and $m$. Then, it is clear that the following
strategies for any small enough positive $\epsilon$ are also
optimal:

$$
\tilde{T}^{a*}_i=
  \begin{cases}
    {T}^{a*}_i & \text{for } i\not=k,m, \\
    {T}^{a*}_i+\epsilon & \text{for } i=k,\\
    {T}^{a*}_i-\epsilon\pi_k/\pi_m & \text{for } i=m,
  \end{cases}
$$
$$
\tilde{T}^{b*}_i=
  \begin{cases}
    {T}^{b*}_i & \text{for } i\not=k,m, \\
    {T}^{b*}_i-\epsilon & \text{for } i=k,\\
    {T}^{b*}_i+\epsilon\pi_k/\pi_m & \text{for } i=m.
  \end{cases}
$$
This completes the proof of Proposition~\ref{rmk1}.

%

\section{A recursive approach to the symmetric water filling game}

Let $\omega^1$,\ldots, $\omega^L$ be some parameters which in the
future will act as Lagrangian multiplies. Using these parameters we
introduce some auxiliary notations. Assume that these parameters are
arranged as follows (this assumption does not reduce the generality
of our forthcoming conclusions):
\begin{equation}
\label{eqnOm} \omega^1\leq\ldots\leq \omega^L.
\end{equation}
Also denote
$$
\bar\omega=(\omega^1,\ldots,\omega^L).
$$
Introduce  the following auxiliary sequence:
$$
t^r=\frac{1}{1-g}\left(\frac{1+(r-1)g}{\omega^{r}}-g\sum_{j=1}^{r}\frac{1}{\omega^j}\right)
\mbox{ for } r\in [1,L].
$$

\noindent It is clear that by (\ref{eqnOm})
$$
t^{r+1}=\frac{1+(r-1)g}{1-g}\left(\frac{1}{\omega^{r+1}}-\frac{1}{\omega^r}\right)+t^r\leq
t^r.
$$
Thus,
$$
t^L\leq t^{L-1}\leq \ldots \leq t^1,
$$
and
\begin{equation}
\begin{split}
\label{eqnDD}
\frac{1}{\omega^{r+1}}-\frac{1}{\omega^{r}}=\frac{1-g}{1+(r-1)g}(t^{r+1}-t^r).
\end{split}
\end{equation}
Hence, for $j\in [k+1,L]$ we have:
\begin{equation}
\begin{split}
\label{eqnDD00}
\frac{1}{\omega^{k}}-\frac{1}{\omega^{j}}=\sum_{r=k}^{j-1}\frac{1-g}{1+(r-1)g}(t^{r}-t^{r+1}).
\end{split}
\end{equation}

Then, sequences $\{\omega^{r}\}$ and $\{t^{r}\}$ has the following
recurrent relations:
\begin{equation}
\label{eqnMult}
\begin{split}
\frac{1}{\omega^{1}}&=t^1,\quad
\frac{1}{\omega^{2}}=(1-g)t^2+gt^1,\\
\frac{1}{\omega^{r+1}}&=\frac{1-g}{1+(r-1)g}t^{r+1}
+\sum^r_{j=2}\frac{(1-g)g}{(1+(j-1)g)(1+(j-2)g)}t^{j}+t^1,
\end{split}
\end{equation}
where $r\ge 1$. If we know sequence $\{t^{r}\}$ we can restore
sequence $\{\omega^{r}\}$. Thus, these two sequences are equivalent.

Introduce one more auxiliary sequence as follows:
$$
\tau^k_r=\frac{1}{1-g}\left(\frac{1+(L-1-r+k)g}{\omega^k}-g\sum_{j=1}^{L-r+k}\frac{1}{\omega^j}\right),
$$
where $r\in [k,L], k\in [1,L]$. There is a simple relation between
sequences $\{\omega^{k}\}$ and $\{t^{k}\}$ and $\{\tau^k_{r}\}$:
\begin{equation}
\label{eqnOmOm0} \tau^k_L=t^k,
\end{equation}
and
\begin{equation}
\label{eqnDD1} \tau^k_r =
\frac{1+(L-1-r+k)g}{1-g}\left(\frac{1}{\omega^k}-\frac{1}{\omega^{L-r+k}}\right)+t^{L-r+k}.
\end{equation}
So, by (\ref{eqnDD00}), collecting terms which depends on $t^k$ we
obtain

\begin{equation}
\label{eqnOmOm} \tau^k_r=b^{k,r}t^k+A^{k,r},
\end{equation}
where
$$
b^{k,r}=\frac{1+(L-1-r+k)g}{1+(k-1)g},
$$
and
\begin{equation*}
\begin{split}
A^{k,r} &=g\sum_{j=k+1}^{L-r+k-1}
\frac{1+(L-1-r+k)g}{(1+(j-1)g)(1+jg)}t^j
-\frac{g}{(1+(L-2-r+k)g}t^{L-r+k}.
\end{split}
\end{equation*}

\noindent Thus, $A^{k,r}$ depends only on $\{t^j\}$ with $j>k$.

\noindent Finally introduce the following notation:

\noindent (a) for $N^0_i < t^L$
$$
T^k_i(\bar\omega)=\frac{1}{1+(L-1)g}(\tau^k_k-N^0_i),
$$

\noindent (b) $t^{L+k+1-r}\leq N^0_i<t^{L+k-r}$ where $r\in [k+1,L]$
$$
T^k_i(\bar\omega)=\frac{1}{1+(L-1-r+k)g}(\tau^k_r-N^0_i),
$$

\noindent (c) for $t^k \leq N^0_i$
$$
T^k_i(\bar\omega)=0.
$$

For others combinations of relations between $\omega^j$, $j\in
[1,L]$, $T^k_i$ are defined by symmetry. By Theorem~\ref{thm02} we
have the following result.

\begin{Theorem}
\label{thmFormOfNE} Each Nash equilibrium is of the form
$(T^1(\bar\omega), \ldots,T^L(\bar\omega))$.
\end{Theorem}

The next lemma provides a nice relation between $L$ and $L-1$ person
games which shows that the introduction of a new user into the game
leads to a bigger competition for the better quality channels
meanwhile users prefer to keep the old structure of their strategies
for worse quality channels.

\begin{Lemma}  \label{lmnMonoton}
Let $(T^{1,L}(\omega_1,\ldots, \omega_L),\ldots,
T^{L,L}(\omega_1,\ldots, \omega_L))$ given by
Theorem~\ref{thmFormOfNE} (here we added the second super-script
index in the notation of the strategies in order to emphasize that
the strategies depend on the number of users). Then, we have
$$
T^{k,L}_i(\omega_1,\ldots, \omega_L)=
  \begin{cases}
    \frac{\displaystyle \tau^k_k-N^0_i}{\displaystyle 1+(L-1)g} & \text{for } N^0_i < t^L, \\
    T^{k,L-1}_i(\omega_1,\ldots, \omega_{L-1}) & \text{for } t^L\leq N^0_i,
  \end{cases}
$$
where $k\in [1,L-1]$ and
$$
T^{L,L}_i(\omega_1,\ldots, \omega_L)=
  \begin{cases}
   \frac{\displaystyle t^L-N^0_i}{\displaystyle 1+(L-1)g}& \text{for } N^0_i < t^L, \\
    0 & \text{for } t^L\leq N^0_i.
  \end{cases}
$$
\end{Lemma}

\section{A water-filling algorithm}

In this section we describe a version of the water-filling algorithm
for finding the NE and supply a simple proof of its convergence
based on some monotonicity properties.

\noindent Let
$$
H^k(\bar\omega)=\sum_{i=1}^n\pi_iT^k_i(\bar\omega) \mbox{ for } k\in
[1,L].
$$
To find a NE we have to find $\bar\omega$ such that
\begin{equation}
\label{eqnFFF} H^k(\bar\omega)= \bar T^k \mbox{ for } k\in [1,L].
\end{equation}

It is clear that $H^k(\bar\omega)$ has the following properties,
collected in the next Lemma, which follow directly from the explicit
formulas of the NE.
\begin{Lemma}
\label{lmnP}(i) $H^k(\bar\omega)$ is nonnegative and continuous,
(ii) $H^k(\bar\omega)$ is decreasing on $\omega^k$, (iii)
$H^k(\bar\omega)\to \infty$ for $\omega^k\to 0$, (iv)
$H^k(\bar\omega)= 0$ for enough big $\omega^k$, say for
$\omega^k\geq 1/N^0_1$, (v) $H^k(\bar\omega)$ is non-increasing by
$\omega^j$ where $j\not= k$.
\end{Lemma}

This properties give a simple proof of the convergence of the
following iterative water filling algorithm for finding the NE.

Let $\omega^k_0$ for all $k\in [1,L]$ be such that
$H^k(\bar\omega_0)= 0$, for example $\omega^k_0=1/N^0_1$. Let
$\omega^k_1=\omega^k_0$ for all $k\in [2,L]$ and define $\omega^1_1$
such that $H^1(\bar\omega_1)= \bar T^1$. Such $\omega^1_1$ exists by
Lemma~\ref{lmnP}(i)-(iii). Then, by Lemma~\ref{lmnP}(i),(v)
$H^k(\bar\omega_0)= 0$ for $k\in [2,L]$. Let $\omega^k_2=\omega^k_1$
for all $k\not=2$ and define $\omega^2_2$ such that
$H^2(\bar\omega_2)= \bar T^2$. Then, by Lemma~\ref{lmnP}(i),(v)
$H^k(\bar\omega_0)= 0$ for $k>2$ and $H^k(\bar\omega_0)\leq \bar
T^k$ for $k=1$ and so on. Let $\omega^k_L=\omega^k_{L-1}$ for all
$k\not=L$ and define $\omega^L_L$ such that $H^L(\bar\omega_L)= \bar
T^L$. Then, by Lemma~\ref{lmnP}(i),(v) $H^k(\bar\omega_L)\leq \bar
T^k$ for $k\not = L$ and so on. So we have non-increasing positive
sequence $\omega^k$. Thus, it converges to an $\bar\omega_*$ which
produces a NE.

\section{Existence and uniqueness of the Nash equilibrium}

In this section we will prove existence and uniqueness of the Nash
equilibrium for $L$ person symmetric water-filling game. Our proof
will have constructive character which allows us to produce an
effective algorithm for finding the equilibrium strategies.

First note that there is a monotonous dependence between the
resources the users can apply and Lagrange multipliers.
\begin{Lemma}
\label{col2} Let $(T^1(\bar\omega), \ldots,T^L(\bar\omega))$ be a
Nash equilibrium. If
\begin{equation}
\label{eqnAssumT} \bar{T}^1\geq\ldots\geq\bar{T}^L
\end{equation}
then (\ref{eqnOm}) holds.
\end{Lemma}

{\it Proof.} The result immediately follows from the following
monotonicity property implied by explicit formulas of the Nash
equilibrium, namely, if $\omega^i<\omega^j$ then
$H^i(\bar\omega)>H^j(\bar\omega)$.

Without loss of generality we can assume that (\ref{eqnAssumT})
holds. Thus, by Lemma~\ref{col2}, (\ref{eqnOm}) also holds.

Let  $\bar\omega$ be the positive solution of (\ref{eqnFFF}). Then,
by Lemma~\ref{col2}, the relation (\ref{eqnOm}) holds. To find
$\bar\omega$ we have to solve the system of non-linear equations
(\ref{eqnFFF}). It is quite bulky system and it looks hard to solve.
We will not solve it directly. What we will do we express
$\omega^1$,\ldots $\omega^L$ by $t^1$, \ldots, $t^L$, substitute
these expression into (\ref{eqnFFF}). The transformed system will
have a triangular form, namely
\begin{equation}
\begin{split}
\label{eqnFFFFFF}
&\tilde{H}^L(t^L)=\bar{T}^L,\\
&\tilde{H}^{L-1}(t^{L-1},t^L)=\bar{T}^{L-1},\\
& \cdots\\
&\tilde{H}^{1}(t^{1},\ldots,t^{L-1}, t^L)=\bar{T}^{1}.\\
\end{split}
\end{equation}
The last system, because of monotonicity properties of $\tilde{H}^k$
on $t^k$, can be easily solved. Now we can move on to construction
of $\tilde{H}^L(t^L)$, \ldots,
$\tilde{H}^1(t^{1},\ldots,t^{L-1},t^L)$. First we will construct
$\tilde{H}^L(t^L)$ and find the optimal $t^L$. Note that,

\begin{equation*}
\begin{split}
H^L(\bar\omega)&= \sum_{N^0_i<t^L} \pi_i T^L_i(\bar\omega)= \\
&=
\frac{1}{1+(L-1)g}\sum_{N^0_i<t^L}\pi_i(\tau^L_L-N^0_i)\\
&=
\frac{1}{1+(L-1)g}\sum_{N^0_i<t^L}\pi_i(t^L-N^0_i)=\tilde{H}^L(t^L).
\end{split}
\end{equation*}
It is clear that $\tilde{H}^L(\cdot)$ is continuous in $(0,\infty)$,
$\tilde{H}^L(\tau)=0$ for $\tau\leq N^0_1$,
$\tilde{H}^L(+\infty)=+\infty$ and  $\tilde{H}^L(\cdot)$ is strictly
increasing in $(N^0_1,\infty)$. Then, there is the unique positive
$t^L_*$ such that
\begin{equation}
\label{eqnTauL} \tilde{H}^L(t^L_*)=\bar{T^L}.
\end{equation}
Now we move on to construction of  $\tilde{H}^{L-1}(t^{L-1},t^L)$
and finding the optimal $t^{L-1}$. Note that
$\tau^{L-1}_{L}=t^{L-1}$ and by (\ref{eqnDD00}) and (\ref{eqnDD1}),
we have
\begin{equation*}
\begin{split}
&\tau^{L-1}_{L-1}=\tau^{L-1}_{L}+\frac{g}{1-g}\left(\frac{1}{\omega^{L-1}}-\frac{1}{\omega^{L}}\right)\\
&=t^{L-1}+\frac{g}{1+(L-2)g}(t^{L-1}-t^{L})\\
&=\frac{1+(L-1)g}{1+(L-2)g}t^{L-1}-\frac{g}{1+(L-2)g}t^{L}.
\end{split}
\end{equation*}
\noindent Thus,
\begin{equation*}
\begin{split}
&H^{L-1}(\bar\omega)= \sum_{N^0_i<t^L} \pi_i T^{L-1}_i(\bar\omega)+
\sum_{t^L\leq N^0_i<t^{L-1}} \pi_i T^{L-1}_i(\bar\omega)= \\
&= \frac{1}{1+(L-1)g}\sum_{N^0_i<t^L}\pi_i(\tau^{L-1}_{L-1}-N^0_i)
+\frac{1}{1+(L-2)g}\sum_{t^L\leq N^0_i<t^{L-1}
}\pi_i(\tau^{L-1}_{L}-N^0_i)\\
&=
\frac{1}{1+(L-1)g}\sum_{N^0_i<t^L}\pi_i\Bigl(\frac{1+(L-1)g}{1+(L-2)g}t^{L-1}-\frac{g}{1+(L-2)g}t^{L}-N^0_i\Bigr)\\
&+\frac{1}{1+(L-2)g}\sum_{t^L\leq N^0_i<t^{L-1}
}\pi_i(t^{L-1}-N^0_i)\\
&=\tilde{H}^{L-1}(t^{L-1},t^L).
\end{split}
\end{equation*}
It is clear that $\tilde{H}^{L-1}(\cdot,t^L_*)$ is continuous and
increasing in $(t^L_*,\infty)$,
$\tilde{H}^{L-1}(\infty,t^L_*)=+\infty$ and
$\tilde{H}^{L-1}(t^L_*,t^L_*)=\tilde{H}^{L}(t^L_*)=\bar{T^L}\leq\bar{T^{L-1}}$.
So, there is the unique positive $t^{L-1}_*$ such that
\begin{equation}
\label{eqnOmegaL-1} \tilde{H}^{L-1}(t^{L-1}_*,t^{L}_*)=\bar{T^L}.
\end{equation}
Next we construct $\tilde{H}^{k}(t^{k},\ldots, t^{L-1},t^L)$ and
find the optimal $t^{k}$ where $k\in [1,L-2]$. By (\ref{eqnOmOm0})
and (\ref{eqnOmOm0}), we have
\begin{equation*}
\begin{split}
&H^{k}(\bar\omega)= \sum_{N^0_i<t^L} \pi_i T^{k}_i+
\sum_{r=k+1}^L\sum_{t^{L+k+1-r}\leq N^0_i<t^{L+k-r}} \pi_i T^{k}_i \\
&=\frac{1}{1+(L-1)g}\sum_{N^0_i<t^L}\pi_i(\tau^k_L-N^0_i)
+\sum_{r=k+1}^L\sum_{t^{L+k+1-r}\leq N^0_i<t^{L+k-r}}\frac{\pi_i(\tau^k_r-N^0_i)}{1+(L-1-r+k)g}\\
&=\frac{1}{1+(L-1)g}\sum_{N^0_i<t^L}\pi_i(b^{k,k}t^k+A^{k,k}-N^0_i)\\
&+\sum_{r=k+1}^L\sum_{t^{L+k+1-r}\leq N^0_i<t^{L+k-r}}\frac{\pi_i(b^{k,r}t^k+A^{k,r}-N^0_i)}{1+(L-1-r+k)g}\\
&=\tilde{H}^{k}(t^{k},t^{k+1},\ldots,t^L).
\end{split}
\end{equation*}
It is clear that $\tilde{H}^{k}(\cdot,t^{k+1}_*, \ldots ,t^L_*)$ is
continuous and increasing in $(t^{k+1}_*,\infty)$,
$\tilde{H}^{k}(\infty,t^{k+1}_*, \ldots, t^L_*)=+\infty$ and by
Lemma~\ref{lmnMonoton} $\tilde{H}^{k}(t^{k+1}_*,t^{k+1}_*,\ldots,
t^L_*)=\tilde{H}^{k+1}(t^{k+1}_*,\ldots,
t^L_*)=\bar{T}^{k+1}\leq\bar{T}^{k}$. So, there is the unique
positive $t^{k}_*$ such that
\begin{equation}
\label{eqnOmegaK}
\tilde{H}^{k}(t^{k}_*,t^{k+1}_*,\ldots,t^{L}_*)=\bar{T^k}.
\end{equation}
Thus, we have proved the following result:
\begin{Lemma}
\label{lmn51} Solution of the system (\ref{eqnFFF}) is equivalent to
solution of the triangular system (\ref{eqnFFFFFF}). This system has
the unique solution which can be found sequentially from $t^L$ down
to $t^1$, applying either the bisection method or the explicit
scheme suggested in Section~II. The optimal Lagrangian multipliers
can be reconstructed from $\{t^r\}$ by (\ref{eqnMult}).
\end{Lemma}

Finally we also have the following result:
\begin{Theorem} \label{thm41}
The symmetric water filling game has the unique Nash equilibrium
$(T^1(\bar\omega_*),\ldots, T^L(\bar\omega_*))$, where
$\bar\omega_*$ is given by Lemma~\ref{lmn51}.
\end{Theorem}

Note that although the payoffs have symmetric form, the equilibrium
strategies, because of triangular form of system (\ref{eqnFFFFFF}),
have hierarchical structure induced by difference in power levels
available to the users. Namely, the user who has to transmit with
smaller average power consumption, in our case it is user $L$, acts
first. He assigns his optimal strategies as if there is no other
users at all but taking into account the total number of users and
fading channels gains. Then, the turn to act is given to user $L-1$.
He takes into account only the behavior of the user $L$ with smaller
average power consumption than he has, the total number of users and
fading channels gains and so on. The last user who constructs the
equilibrium strategy is user 1 with the largest available power
resource.

\section{Closed form solution for $L$ person game}
\label{sec:GameClosedForm}

In this section for the case of $L$ users we show how
Theorem~\ref{thm41} and Lemma~\ref{lmn51} can be used to construct
NE in closed form.

Assume that $\bar T^1 > \ldots > \bar T^L$. We will construct the
optimal strategies $T^{L*}$, \ldots, $T^{1*}$ sequentially.

{\it Step for construction of $T^{L*}$}. Since $\tilde{H}^L(\cdot)$
is strictly increasing we can find an integer $k_L$ such that
$$
\tilde{H}^L(N^0_{k^L})<\bar{T}^L\leq \tilde{H}^L(N^0_{k^L+1}).
$$
or from the following equivalent conditions:
\begin{equation*}
\label{eqnL} \varphi^L_{k^L}<\bar {T}^L \leq \varphi^L_{k^L+1},
\end{equation*}
where
$$
\varphi^L_k=\frac{1}{1+(L-1)g}\sum_{i=1}^k\pi_i(N^0_k-N^0_i),
$$
for $k \leq n$, and $\varphi^L_{n+1}=\infty$. Then, since
$\tilde{H}^L(t^L_*)=\bar{T}^L$, we have that
\begin{equation*}
\label{eqnTau3}
t^L_*=\frac{(1+(L-1)g)\bar{T}^L+\sum^{k_L}_{i=1}\pi_iN^0_i}{\sum^{k_L}_{i=1}\pi_i}.
\end{equation*}
\noindent Thus, the optimal strategy of user $L$ is given as follows
$$
 T^{L*}_i =
 \begin{cases}
    \frac{\displaystyle  1}{\displaystyle 1+(L-1)g}(t^L_*-N^0_i)& \text{if } i\in [1,k^L], \\
    0 & \text{if }  i\in [k^L+1,n].
  \end{cases}
$$

{\it Step for construction of $T^{(L-1)*}$}. Since $t^{L-1}_*$ is
the root of the equation $\tilde{H}^{L-1}(\cdot, t^L_*)=\bar
{T}^{L-1}$ there is $k^{L-1}$ such that $k^{L-1}\geq k^{L}$ and
$N^0_{k^{L-1}+1}\geq t^{L-1}_*
> N^0_{k^{L-1}}$. Thus,

\begin{equation*}
\begin{split}
\label{eqnL-1} t^{L-1}_*&=\Bigl( \bar{T}^{L-1}+
\frac{1}{1+(L-2)g}\sum_{i={k^L+1}}^{k^{L-1}}\pi_iN^0_i\\
&+\frac{1}{1+(L-1)g}\sum_{i=1}^{k^L}\pi_i(\frac{gt^*_{L}}{1+(L-2)g}+N^0_i)\Bigr)\\
&\Bigr/\Bigl(\frac{1}{1+(L-2)g} \sum_{i=1}^{k^{L-1}}\pi_i\Bigr).
\end{split}
\end{equation*}
Here and bellow we assume that $\sum_x^y 1 = 0$ for $y<x$. So,
$k^{L-1}\geq k^L$ can be found as follows:

(i) $k^{L-1}=k^{L}$ if $\bar {T}^{L-1} \leq
\varphi^{L-1}_{k^{L-1}+1}$,

(ii) otherwise $k^{L-1}$ is given by the condition:
\begin{equation*}
\label{eqnL-12} \varphi^{L-1}_{k^{L-1}}<\bar {T}^{L-1} \leq
\varphi^{L-1}_{k^{L-1}+1},
\end{equation*}
where
\begin{equation*}
\begin{split}
\varphi^{L-1}_{k}&=\sum_{i=k^{L}+1}^{k}\frac{\pi_i}{1+(L-2)g}(N^0_{k}-N^0_i)\\&
+\sum_{i=1}^{k^{L}}\frac{\pi_i}{1+(L-1)g}\\
&\times\left(\frac{1+(L-1)g}{1+(L-2)g}N^0_{k}-N^0_i-\frac{g}{1+g}t^{L-1}_*\right),
\end{split}
\end{equation*}
for $k \in [k^{L-1}+1,n]$ and $\varphi^{L-1}_{n+1}=\infty$.

\noindent Thus, the optimal strategy $T^{(L-1)*}$ of user $L-1$ is
given by
$$
T^{(L-1)*}_i =
 \begin{cases}
    \frac{\displaystyle t^{L-1}_*}{\displaystyle 1+(L-2)g}\\
    -\frac{\displaystyle \frac{g}{1+g}t^L_*+N^0_i}{\displaystyle
    1+(L-1)g},
     &  i \in [1,k^L], \\
    \frac{\displaystyle  1}{\displaystyle 1+(L-2)g}(t^{L-1}_*-N^0_i), & i \in [k^L+1,k^{L-1}],\\
    0, & i \in [k^{L-1}+1,n].
  \end{cases}
$$

{\it Step for construction of $T^{M*}$ where $M<L$}. We have already
constructed $T^{L*}$, \ldots, $T^{(M+1)*}$ and now we are going to
construct $T^{M*}$. Since $t^{M}_*$ is the root of the equation
$\tilde{H}^{M}(\cdot, t^{M+1}_*,\ldots, t^L_*)=\bar {T}^{M}$ there
is $k^{M}$ such that $k^{M}\geq k^{M+1}$ and $N^0_{k^{M}+1}\geq
t^{M}_* > N^0_{k^{M}}$. Thus,

\begin{equation*}
\begin{split}
\label{eqnM}
t^{M}_*&=\Bigl(\bar{T}^{M}+\frac{1}{1+(L-1)g}\sum_{i=1}^{k_M}\pi_i(A^{k,k}-N^0_i)\\
&+\sum_{r=M+1}^L\sum_{i=k^p+1}^{k^{p-1}}\frac{\pi_i(A^{p,r}-N^0_i)}{1+(L-1-r+p)g})\\
&\Bigr/\Bigl(\frac{1}{1+(M-1)g} \sum_{i=1}^{k^{M}}\pi_i\Bigr).
\end{split}
\end{equation*}
So, $k^{M}\geq k^{M+1}$ can be found as follows:

(i) $k^{M}=k^{M+1}$ if $\bar {T}^{M} \leq \varphi^{M}_{k^{M}+1}$,

(ii) otherwise $k^{M}$ is given by the condition:
\begin{equation*}
\label{eqnM2} \varphi^{M}_{k^{M}}<\bar {T}^{M} \leq
\varphi^{M}_{k^{M}+1}
\end{equation*}
where
\begin{equation*}
\begin{split}
\varphi^{M}_{k}
&=\frac{1}{1+(L-1)g}\sum_{i=1}^k\pi_i(b^{k,k}N^0_k+A^{k,k}-N^0_i)\\
&+\sum_{r=M+1}^L\sum_{i={k^p+1}}^{k^{p-1}}\frac{\pi_i(b^{p,r}N^0_k+A^{p,r}-N^0_i)}{1+(L-1-r+p)g}.
\end{split}
\end{equation*}
\noindent Thus, the optimal strategy of user $L$ is given as follows
$$
T^{M*}_i =
\begin{cases}
    \frac{\displaystyle \tau^M_M-N^0_i}{\displaystyle 1+(L-1)g},
     &  i \in [1,k^L], \\
    \frac{\displaystyle \tau^M_r-N^0_i}{\displaystyle 1+(L-1-r+M)g}, & i \in [k^r+1,k^{r-1}], \\
    &r\in [M+1,L]\\
    0, & i \in [k^{M}+1,n].
  \end{cases}
$$
In particular for two and three person games ($L=2$ and $L=3$) we
have the following results.
\begin{Theorem}
\label{thm2player} Let $\bar T_1> \bar T_2$. Then, the Nash
equilibrium strategies are given by
    \begin{equation*}
    \label{eqn2P}
\begin{split}
T^{1*}_i &=
  \begin{cases}
    t^1_*-\frac{\displaystyle g t^2_*+N^0_i}{\displaystyle 1+g}
     & \text{if } i \in [1,k^2], \\
    t^1_*-N^0_i & \text{if }  i \in [k^2+1,k^1],\\
    0 & \text{if }  i \in [k^1+1,n],
  \end{cases}\\
T^{2*}_i &=
 \begin{cases}
    \frac{\displaystyle  1}{\displaystyle 1+g}(t^2_*-N^0_i)& \text{if } i\in [1,k^2], \\
    0 & \text{if }  i\in [k^2+1,n],
  \end{cases}\\
\end{split}
\end{equation*}
where

(a) $k^2$, $t^2_*$  are given by
\begin{equation*}
\label{eqnTau2}
t^2_*=\frac{(1+g)\bar{T}^2+\sum^{k_2}_{i=1}\pi_iN^0_i}{\sum^{k_2}_{i=1}\pi_i},
\end{equation*}
$k_2$ can be found from the condition
$$
\varphi^2_{k^2}<\bar {T}^2 \leq \varphi^2_{k^2+1},
$$
where
$$
\varphi^2_k=\frac{1}{1+g}\sum_{i=1}^k\pi_i(N^0_k-N^0_i),
$$
for $k \leq n$, and $\varphi^2_{n+1}=\infty$,

(b) $k^1$ and $t^1_*$  are given by
\begin{equation*}
\label{eqnOmega2} t^1_*=\frac{\displaystyle \bar{T}^1+
\sum_{i={k^2+1}}^{k^1}\pi_iN^0_i+\frac{1}{1+g}\sum_{i=1}^{k^2}\pi_i(gt^*_2+N^0_i)}{\displaystyle
\sum_{i=1}^{k^1}\pi_i},
\end{equation*}
$k^1\geq k^2$ can be found as follows:

(i) $k^1=k^2$ if $\bar {T}^1 \leq  \varphi^1_{k^2+1}$

(ii) otherwise $k^1$ is given by the condition:
\begin{equation*}
\label{eqn2p2} \varphi^1_{k^1}<\bar {T}^1 \leq  \varphi^1_{k^1+1},
\end{equation*}
where
\begin{equation*}
\begin{split}
\varphi^1_{k}&=\sum_{i=k^2+1}^{k}\pi_i(N^0_{k}-N^0_i)\\&
+\frac{1}{1+g}\sum_{i=1}^{k^2}\pi_i\left((1+g)N^0_{k}-N^0_i-gt^2_*\right)
\end{split}
\end{equation*}
for $k \in [k_2+1,n]$, and $\varphi^1_{n+1}=\infty$.
\end{Theorem}

\begin{Theorem}
\label{thm3player} Let $\bar T_1> \bar T_2> \bar T_3$. Then, the
Nash equilibrium strategies are given by
\begin{equation*}
\begin{split}
T^{1*}_i &=
  \begin{cases}
    t^1_*-
\frac{\displaystyle g t^2_*}{\displaystyle 1+g}
    -\frac{\displaystyle \frac{g t^3_*}{1+g}+N^0_i}{\displaystyle 1+2g}
     & \text{if } i \in [1,k^3],\\
    t^1_*-\frac{\displaystyle g t^2_*+N^0_i}{\displaystyle 1+g}
     & \text{if } i \in [k^3+1,k^2], \\
    t^1_*-N^0_i & \text{if }  i \in [k^2+1,k^1],\\
    0 & \text{if }  i \in [k^1+1,n],
  \end{cases}\\
T^{2*}_i &=
 \begin{cases}
    \frac{\displaystyle t^2_*}{\displaystyle 1+g}-\frac{\displaystyle \frac{g}{1+g}t^3_*+N^0_i}{\displaystyle 1+2g}
     & \text{if } i \in [1,k^3], \\
    \frac{\displaystyle  1}{\displaystyle 1+g}(t^2_*-N^0_i) & \text{if }  i \in [k^3+1,k^2],\\
    0 & \text{if }  i \in [k^2+1,n],
  \end{cases}\\
 T^{3*}_i &=
 \begin{cases}
    \frac{\displaystyle  1}{\displaystyle 1+2g}(t^3_*-N^0_i)& \text{if } i\in [1,k^3], \\
    0 & \text{if }  i\in [k^3+1,n],
  \end{cases}\\
\end{split}
\end{equation*}
where

(a) $k^3$, $t^3_*$ are given by
\begin{equation*}
\label{eqnTau33}
t^3_*=((1+2g)\bar{T}^3+\sum^{k_3}_{i=1}\pi_iN^0_i)/(\sum^{k_3}_{i=1}\pi_i),
\end{equation*}
\begin{equation*}
\label{eqn3p1} \varphi^3_{k^3}<\bar {T}^3 \leq \varphi^3_{k^3+1},
\end{equation*}
and
$$
\varphi^3_k=\frac{1}{1+2g}\sum_{i=1}^k\pi_i(N^0_k-N^0_i),
$$
for $k \leq n$, and $\varphi^3_{n+1}=\infty$,

(b) $k^2$, $t^2_*$ are given by
\begin{equation*}
\begin{split}
\label{eqnOmega3} t^2_*&=\Bigl( \bar{T}^2+
\frac{1}{1+g}\sum_{i={k^3+1}}^{k^2}\pi_iN^0_i
+\frac{1}{1+2g}\sum_{i=1}^{k^3}\pi_i(\frac{gt^3_*}{1+g}+N^0_i)\Bigr)\Bigr/\Bigl(\frac{1}{1+g}
\sum_{i=1}^{k^2}\pi_i\Bigr),
\end{split}
\end{equation*}
(i) $k^2=k^3$ if $\bar {T}^2 \leq  \varphi^2_{k^3+1}$,

(ii) otherwise $k^2$ is given by the condition:
\begin{equation*}
\label{eqn3p2} \varphi^2_{k^2}<\bar {T}^2 \leq  \varphi^2_{k^2+1}
\end{equation*}
and
\begin{equation*}
\begin{split}
\varphi^2_{k}&=\sum_{i=k^3+1}^{k}\frac{\pi_i}{1+g}(N^0_{k}-N^0_i)
+\sum_{i=1}^{k^3}\pi_i\left(\frac{1}{1+g}N^0_{k}-\frac{N^0_i+gt^3_*/(1+g)}{1+2g}\right).
\end{split}
\end{equation*}
for $k \in [k^3+1,n]$ and $\varphi^2_{n+1}=\infty$

(c) $k^1$, $t^1_*$ are given by
\begin{equation*}
\begin{split}
\label{eqnOmega33} t^1_*=&\Bigl(\displaystyle \bar{T}^1+
\sum_{i={k^2+1}}^{k^1}\pi_iN^0_i+\sum_{i=k^3+1}^{k^2}\pi_i\frac{\displaystyle
g t^2_*+N^0_i}{\displaystyle 1+g}
+\sum_{i=1}^{k^3}\pi_i\Bigl(\frac{\displaystyle g
t^2_*}{\displaystyle 1+g}
    +\frac{\displaystyle \frac{g t^3_*}{1+g}+N^0_i}{\displaystyle 1+2g}\Bigr)\Bigr)\Bigr/
\sum_{i=1}^{k^1}\pi_i.
\end{split}
\end{equation*}
So, $k^1\geq k^2$ can be found as follows:

(i) $k^1=k^2$ if $\bar {T}^1 \leq  \varphi^1_{k^2+1}$,

(ii) otherwise $k^1$ is given by the condition:

\begin{equation*}
\label{eqn3p22} \varphi^1_{k^1}<\bar {T}^1 \leq  \varphi^1_{k^1+1}
\end{equation*}
where
\begin{equation*}
\begin{split}
 \varphi^1_{k}&=\sum_{i={k^2+1}}^{k}\pi_i(N^0_k-N^0_i)+\sum_{i=k^3+1}^{k^2}\pi_i\bigl(N^0_k-\frac{\displaystyle g t^2_*+N^0_i}{\displaystyle 1+g}\bigr)
+\sum_{i=1}^{k^3}\pi_i\Bigl(N^0_k- \frac{\displaystyle g
t^2_*}{\displaystyle 1+g}
    -\frac{\displaystyle \frac{g t^3_*}{1+g}+N^0_i}{\displaystyle
    1+2g}\Bigr).
\end{split}
\end{equation*}

\end{Theorem}

\section{Numerical examples}
\label{sec:numeric}

Let us demonstrate the closed form approach by  numerical examples.
Take $n=5$, $N^0_i=\kappa^{i-1}$, $\kappa=1.7$, $\pi_i=1/5$ for
$i\in [1,5]$. We consider the cases 1, 2 and 3 users scenari.

{\it Single user scenario}.  Let $\bar T=5$. Then, by
Theorem~\ref{thm1player} as the first step we calculate $\varphi_t$
for $t\in [1,5]$. In our case we get (0, 0.14, 0.616, 1.8298,
4.58108). Thus, we have $k=5$ and the optimal water-filling strategy
is $T^*=(7.771, 7.071, 5.881, 3.858, 0.419)$ with payoff 1.11.

{\it Two users scenario}. Let also $g=0.9$, $\bar T^1=5$, $\bar
T^2=1$. Then, by Theorem~\ref{thm2player} as the first step we
calculate $\varphi^2_t$ for $t\in [1,5]$. In our case we get (0,
0.074, 0.324, 0.963, 2.411). Thus, $k^2=4$ and $t^2_*=5.001$. Then
we calculate $\varphi^1_t$ for $t=5$. In our case we get 6.994052.
Thus, $k^1=4$ and $t^1_*=0.010$. Therefore, we have the following
equilibrium strategies $T^{1*}=(7.106, 6.737, 6.111, 5.046, 0)$ and
$T^{2*}=(2.106, 1.737, 1,111, 0.0462, 0)$ with payoffs 0.801 and
0.116, respectively.

{\it Three users scenario}. Let us introduce the third player with
the average power constraint $\bar T^3=0.5$. Then, by
Theorem~\ref{thm3player} we can find that $T^{1*}=(6.419, 6.169,
5.744, 4.900, 1.769)$, $T^{2*}=(1.861, 1.611, 1.186, 0.342, 0)$ and
$T^{3*}=(1.142, 0.892, 0.467, 0, 0)$ are equilibrium strategies with
payoffs 0.728, 0.113 and 0.055, respectively.

\begin{figure}[h]
  \begin{minipage}{1.0\textwidth}
    \begin{center}
        \centering {\epsfxsize=3.3in \epsfbox{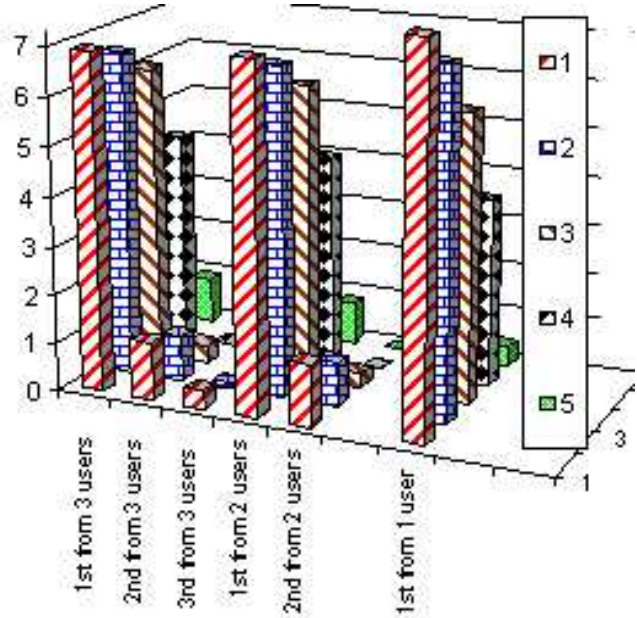}}
        \caption{Optimal strategies for 1, 2 and 3 user games }
        \label{fig:3players}
    \end{center}
   \end{minipage}
  \end{figure}

The equilibrium strategies of all three cases are shown in
Figure~\ref{fig:3players}. When a new user comes into competition,
it leads to a bigger rivalry for using good quality channels and it
results in the situation when bad quality channels turn out to
become more attractive for users than they were when there were
smaller number of users.

\begin{figure}[h]
  \begin{minipage}{1.0\textwidth}
    \begin{center}
        \centering {\epsfxsize=3.6in \epsfbox{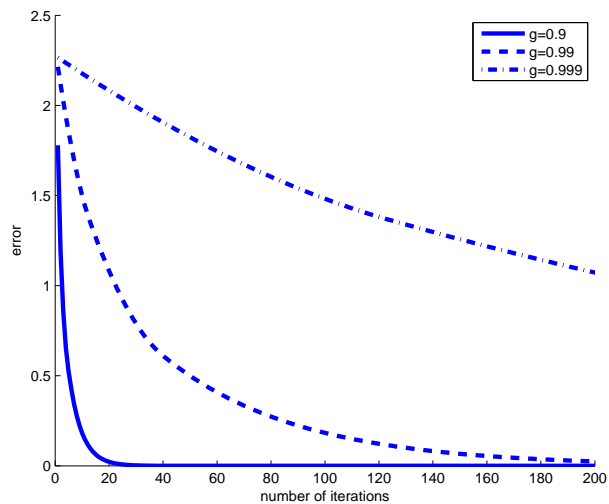}}
        \caption{Convergence of IWFA}
        \label{fig:IWFA}
    \end{center}
   \end{minipage}
  \end{figure}

We have run IWFA, which produced the same values for the optimal
strategies and payoffs. However, we have observed that the
convergence of IWFA is slow when $g \approx 1$. In
Figure~\ref{fig:IWFA}, for the two users scenario, we have plotted
the total error in strategies
$||T^1_k-T^{1*}||_2+||T^2_k-T^{2*}||_2$, where $T^i_k$ are the
strategies produced by IWFA on the $k$-th iteration and $T^{i*}$ are
the Nash equilibrium strategies. Our approach instantaneously finds
the Nash equilibrium for all values of $g$. Also, it is interesting
to note that by Theorems~\ref{thm2player} and \ref{thm3player} the
quantity of channels as well as the channels themselves used by
weaker user (with smaller resources) is independent from the
behavior of the stronger user (with larger resources). Of course,
each user allocates his/her resources among the channels taking into
account the opponent behavior.

\begin{figure}[h]
  \begin{minipage}{1.0\textwidth}
    \begin{center}
        \centering {\epsfxsize=3.6in \epsfbox{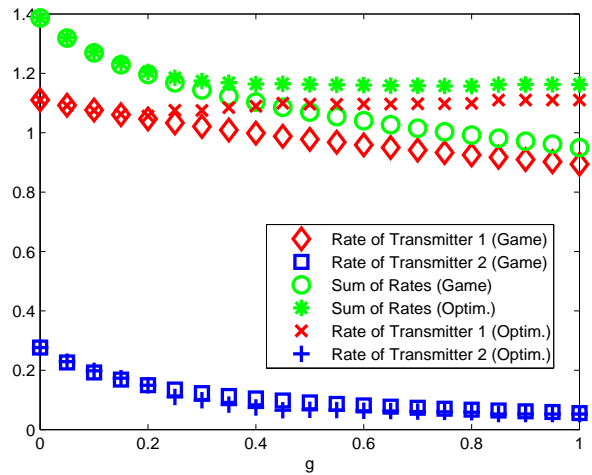}}
        \caption{Centralized Optimization vs. Game}
        \label{fig:OptvsGame}
    \end{center}
   \end{minipage}
  \end{figure}

\begin{figure}[h]
  \begin{minipage}{1.0\textwidth}
    \begin{center}
        \centering {\epsfxsize=3.6in \epsfbox{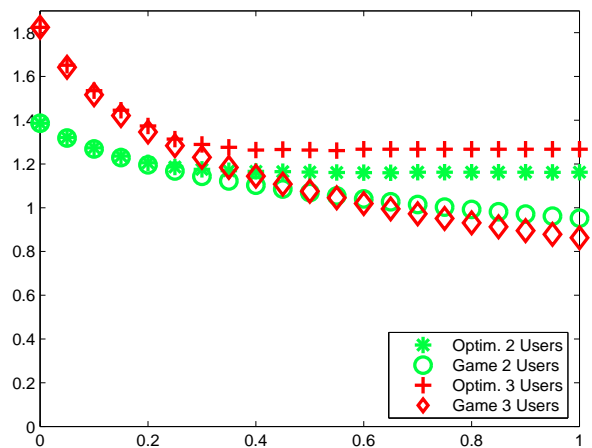}}
        \caption{The effect of a new user}
        \label{fig:OptvsGame23}
    \end{center}
   \end{minipage}
  \end{figure}

In Figures~\ref{fig:OptvsGame} and \ref{fig:OptvsGame23}, we compare
the non-cooperative approach with the cooperative approach.
Specifically, we compare the transmission rates and their sum under
Nash equilibrium strategies and under strategies obtained from the
centralized optimization of the sum of users' rates. The main
conclusions are: the cost of anarchy is nearly zero for $g \in
[0,1/4]$ and then it grows up to 22\% when $g$ grows from $1/4$ to
$1$; the user with more resources gains significantly more from the
centralized optimization. Hence, the non-cooperative approach
results in a more fair resource distribution. In
Figure~\ref{fig:OptvsGame23} we plot the total transmission rate
under Nash equilibrium strategies and under strategies obtained from
the centralized optimization for the cases of 2 and 3 users. As
expected the introduction of a new user increases the cost of
anarchy. Furthermore, in the case of the centralized optimization
with the introduction of a new user the total rate increases, and on
contrary in the game setting the total rate decreases.


\section{Conclusion}
\label{sec:Conc}

We have considered power control for wireless networks in
optimization and game frameworks. Closed form solutions for the
water filling optimization problem and $L$ users symmetric water
filling games have been provided. Namely, now one can calculate
optimal/equilibrium strategies with a finite number of arithmetic
operations. This was possible due to the intrinsic hierarchical
structure induced by the quantity of the resources available to the
users. We have also provided a simple alternative proof of
convergence for a version of iterative water filling algorithm. It
had been known before that the iterative water filling algorithm
converges very slow when the crosstalk coefficient is close to one.
For our closed form approach possible proximity of the crosstalk
coefficient to one is not a problem. We have shown that when the
crosstalk coefficient is equal to one, there is a continuum of Nash
equilibria. Finally, we have demonstrated that the price of anarchy
is small when the crosstalk coefficient is small and that the
decentralized solution is better than the centralized one with
respect to fairness.

\tableofcontents


\begin{thebibliography}{12}

\bibitem{AAG07}
E. Altman, K. Avrachenkov, A. Garnaev, ``A jamming game in wireless
networks with transmission cost''. in {\it Proc. of NET-COOP 2007.
Lecture Notes in Computer Science}, v.4465, pp.1-12, 2007.

\bibitem{AAMP07}
E. Altman, K. Avrachenkov, G. Miller and B. Prabhu, ``Discrete power
control: cooperative and non-cooperative optimization'', in
Proceedings of {\it IEEE INFOCOM 2007}. An extended version is
available as INRIA Research Report no.5818.

\bibitem{CT1991}
T. Cover and J. Thomas, {\it Elements of Information Theory}, Wiley,
1991.

\bibitem{HCB00}
W.~R. Heinzelman, A.~Chandrakasan, and H.~Balakrishnan,
``Energy-efficient
  communication protocol for wireless microsensor networks,'' in \emph{Proc. of
  the 33rd Annual Hawaii International Conference on System Sciences}, v.2,
  Jan. 2000.

\bibitem{G2000}
A. Garnaev, {\it Search Games and Other Applications of Game
Theory}, Springer, 2000.

\bibitem{GV1997}

A.J. Goldsmith and P.P. Varaiya, ``Capacity of fading channels with
channel side information'', {\it IEEE Trans. Information Theory},
v.43(6), pp.1986-1992, 1997.

\bibitem{KG99}
T.~J. Kwon and M.~Gerla, ``Clustering with power control,'' in
\emph{Proc. IEEE Military Communications Conference ({MILCOM}'99)},
v.2, Atlantic City, NJ, USA, 1999, pp.1424--1428.

\bibitem{ElG05} L. Lai and H. El Gamal,
``The water-filling game in fading multiple access channels'',
submitted to {\it IEEE Trans. Information Theory}, 2005.

\bibitem{LG97}
C.~R. Lin and M.~Gerla, ``Adaptive clustering for mobile wireless
networks,'' \emph{{IEEE} JSAC}, v.15, no.7, pp.1265--1275, 1997.

\bibitem{LP06}
Z.-Q. Luo and J.-S. Pang, ``Analysis of iterative waterfilling
algorithm for multiuser power control in digital subscriber lines'',
{\it EURASIP Journal on Applied Signal Processing}, 2006.

\bibitem{PR03}
O. Popescu and C. Rose, ``Water filling may not good neighbors
make'', in {\it Proceedings of GLOBECOM 2003}, v.3, pp.1766--1770,
2003.

\bibitem{PPR04}
D.C. Popescu, O. Popescu and C. Rose, ``Interference avoidance
versus iterative water filling in multiaccess vector channels'', in
{\it Proceedings of IEEE VTC 2004 Fall}, v.3, pp.2058--2062, 2004.

\bibitem{SCGC02} K.B. Song, S.T. Chung, G. Ginis and J.M. Cioffi,
``Dynamic spectrum management for next-generation DSL systems, {\it
IEEE Communications Magazine}, v.40, pp.101--109, 2002.

\bibitem{TV2005}
D. Tse and P. Viswanath, {\it Fundamentals of Wireless
Communication}, Cambridge University Press, 2005.

\bibitem{Yu2002}
W. Yu, {\it Competition and cooperation in multi-user communication
environements}, PhD Thesis, Stanford University, June 2002.

\bibitem{YGC02} W. Yu, G. Ginis and J.M. Cioffi, ``Distributed
multiuser power control for digital subscriber lines'', {\it IEEE
JSAC}, v.20, pp.1105--1115, 2002.

\end{thebibliography}
\end{document}